\documentclass[journal,draftcls,onecolumn]{IEEEtran}
\usepackage[T1]{fontenc}
\usepackage[latin1]{inputenc}
\usepackage{graphicx}
\usepackage{theorem}
\usepackage{amsmath}
\usepackage{amsfonts}
\usepackage{stfloats}
\usepackage{fancyhdr}

\makeatletter

{\theorembodyfont{\rmfamily}
   \newtheorem{The}{{\textbf Theorem}}[section]}
{\theorembodyfont{\rmfamily}
   \newtheorem{Pro}{{\textbf Proposition}}[section]}
{\theorembodyfont{\rmfamily}
   \newtheorem{Lem}{{\textbf Lemma}}[section]}
{\theorembodyfont{\rmfamily}
   }
{\theorembodyfont{\rmfamily}
   \newtheorem{Def}{{\textbf Definition}}[section]}
{\theorembodyfont{\rmfamily}
   \newtheorem{Exa}{{\textbf Example}}[section]}
 
\def\HH{\mathbf{H}}
\def\0{\mathbf{0}}
\def\Hconv{\HH_\mathrm{conv}}

\makeatother
\begin{document}

\pagestyle{fancy}
\fancyhf{} 
\fancyhead[L]{\textit{Accepted for publication in IEEE Communications Letters - Copyright transferred to IEEE}}

\title{Progressive Differences Convolutional\\ Low-Density Parity-Check Codes}

\author{Marco Baldi,~\IEEEmembership{Member,~IEEE}, Marco Bianchi,\\ Giovanni Cancellieri, and Franco Chiaraluce,~\IEEEmembership{Member,~IEEE}

\thanks{M. Baldi, M. Bianchi, G. Cancellieri and F. Chiaraluce are with the Dipartimento di Ingegneria dell'Informazione, Università Politecnica delle Marche, Ancona, Italy (e-mail: \{m.baldi, m.bianchi, g.cancellieri, f.chiaraluce\}@univpm.it).

This work was supported in part by the MIUR project ``ESCAPADE'' (Grant RBFR105NLC) under the ``FIRB - Futuro in Ricerca 2010'' funding program.}}

\maketitle

\thispagestyle{fancy}

\begin{abstract}
We present a new family of low-density parity-check (LDPC) convolutional codes
that can be designed using ordered sets of progressive differences.
We study their properties and define a subset of codes in this
class that have some desirable features, such as fixed minimum distance
and Tanner graphs without short cycles.
The design approach we propose ensures that these properties are guaranteed 
independently of the code rate.
This makes these codes of interest in many practical applications,
particularly when high rate codes are needed for saving bandwidth.
We provide some examples of coded transmission schemes exploiting 
this new class of codes.
\end{abstract}


\section{Introduction}
An increasing interest has been devoted, in recent years, to low-density parity-check
(LDPC) convolutional codes, which have revitalized the field of classical convolutional
coded transmission schemes by exploiting the powerful error correcting performance of 
LDPC codes.

Traditional convolutional codes \cite{Elias1955} are among the most important
classes of channel codes and have found many applications in past and present
communications systems, especially when low latency is needed.
Furthermore, convolutional codes are the main components of turbo codes \cite{Berrou1993},
that deeply changed the scenario of forward error correction.

Convolutional codes are usually designed by optimizing the generator polynomial
and the associated matrix $\mathbf{G}$.
The goal is to keep the number of memory elements in the encoder shift registers
(that is, the constraint length) as small as possible while maximizing the free 
distance \cite{Lin2004Book, Johannesson1999}.
In this way, good codes with limited decoding complexity can be designed.

The advent of LDPC codes \cite{Gallager, Richardson2001} changed this point of view, because 
the computational complexity of LDPC decoding performed over Tanner graphs with belief propagation
algorithms is mostly influenced by the density of symbols $1$ in the 
parity-check matrix $\mathbf{H}$.
Adopting low-density parity-check matrices and avoiding short cycles in the Tanner graph 
ensures to achieve good performance and low decoding complexity.

In the literature, two main methods for designing LDPC convolutional codes have been proposed
\cite{Tanner2004, Felstrom1999} and have originated a number of variants and improvements
(see \cite{Pusane2011} and the references therein).
Both approaches start from LDPC block codes and rearrange their parity-check matrices
through suitable unwrapping procedures to obtain convolutional codes.

In this paper, we present a method to design LDPC convolutional codes
having very simple structured matrices (thus yielding very low encoding and decoding complexity)
and, at the same time, good distance properties and no short cycles in their Tanner graphs.
According to previous approaches, high constraint lengths and time-varying
codes might be required to achieve good performance.
Moreover, the minimum distance of the LDPC convolutional codes so obtained is usually unknown.
On the contrary, our aim is to keep the constraint length as small as possible and
to design time-invariant codes, in order to limit complexity.
In addition, the proposed method allows designing codes with known minimum distance,
which does not depend on the code rate.

The proposed codes are based on ordered sets of progressive differences as separations
between symbols $1$ in their parity-check matrix columns.
For this reason, we call these codes {\itshape progressive differences convolutional} (PDC) LDPC codes.

The organization of the paper is as follows:
in Section \ref{sec:Definition}, we define the new codes and study their properties;
in Section \ref{sec:CodeExamples}, we report some examples of code design according to the proposed approach;
finally, Section \ref{sec:Conclusion} is devoted to concluding remarks.

\section{Definition and properties of PDC-LDPC codes}
\label{sec:Definition}

As for LDPC convolutional codes, each PDC-LDPC code is
defined by a semi-infinite binary parity-check matrix $\Hconv$ \cite{Pusane2011}.
For classical LDPC convolutional codes, $\Hconv$ is an all-zero matrix except 
for a set of binary submatrices $\HH_i(t)$, $t \in \mathbb{Z}_{\ge 0}$,
$i = 0, 1, \ldots, L_h-1$\footnote{Note that $L_h$ measures the memory order
of the syndrome former and it has the same meaning of $m_s + 1$ in \cite{Tanner2004, Pusane2011}.},
which form a staircase from top left towards bottom right within $\Hconv$ \cite{Pusane2011}.
Each matrix $\HH_i(t)$ has size $(a-b) \times a$ and the design rate of the code is, therefore, $R_d = b/a$.
When $t_1 \ne t_2 \Rightarrow \HH_i(t_1) \ne \HH_i(t_2)$, the convolutional LDPC code
is said to be time-varying; otherwise, if $\HH_i(t)$ 
does not depend on $t$, then the code is said to be time-invariant.
The structure of $\Hconv$ for a time-invariant LDPC convolutional code is shown
in \eqref{eq:Hconv}, where $\0$ represents the all-zero $(a-b) \times a$ matrix.

PDC-LDPC codes, here proposed, are a family of time-invariant LDPC
convolutional codes with $b = (a-1)$ and $R_d = (a-1)/a$.
Therefore, each $\HH_i$ is a row vector with length $a$.
A block column formed by $L_h$ submatrices $\HH_0, \HH_1,\ldots, \HH_{L_h-1}$
is denoted by $\HH_s^T$ and coincides with the transpose of the $a \times L_h$ syndrome former matrix, $\HH_s$.
It follows that the syndrome former constraint length is equal to $v_s = L_h \cdot a$.
This value should be kept as small as possible.

\begin{equation}
\Hconv = \left[\begin{array}{cccccc}
\HH_0 			& \0 					& \0 					& \cdots \\
\HH_1				& \HH_0				& \0 					& \cdots \\
\HH_2				& \HH_1				& \HH_0				& \cdots \\
\vdots 			& \vdots 			& \vdots 			& \ddots \\
\HH_{L_h-1}	& \HH_{L_h-2}	& \HH_{L_h-3}	& \ddots \\
\0 					& \HH_{L_h-1}	& \HH_{L_h-2}	& \ddots \\
\0 					& \0 					& \HH_{L_h-1}	& \ddots \\
\0 					& \0 					& \0					& \ddots \\
\vdots			& \vdots			& \vdots			& \ddots \\
\end{array}\right].
\label{eq:Hconv}
\end{equation}

In PDC-LDPC codes, the matrix $\HH_s^T$ has columns with fixed Hamming weight $w$.
The first row is the all-one vector, while the other $L_h-1$
rows have Hamming weight $0$ or $1$.
The matrix $\HH_s^T$ can be defined by $a$ ordered sets of $w-1$ values each.
The $i$-th of these sets is denoted as $D_i = \{d_{i,1}, d_{i,2}, \ldots, d_{i,w-1}\}$, $i = 1,2,\ldots,a$.
Each value $d_{i,j}$, with $i = 1,2,\ldots,a$ and $j = 1,2,\ldots,w-1$, is the difference
between the row positions of the $(j+1)$-th and the $j$-th symbol $1$ within the
$i$-th column of the matrix $\HH_s^T$.
For each ordered set $D_i$, the set $S_i$, $i = 1,2,\ldots,a$, contains all the elements 
in $D_i$ and all the sums of two or more consecutive elements in $D_i$.
We also define the multisets $D = D_1 \uplus D_2 \uplus \ldots \uplus D_a$
and $S = S_1 \uplus S_2 \uplus \ldots \uplus S_a$, where $\uplus$ denotes
the multiset sum.

In the following, we recall some propositions that are known in coding theory
(proofs are omitted for the sake of brevity)
and demonstrate some lemmas and theorems on the properties of PDC-LDPC codes.

\begin{Lem}
\label{lem:CyclesAbsence}
Given $D_1, D_2, \ldots, D_a$ and the corresponding $S_1, S_2, \ldots, S_a$, 
if $S$ does not contain duplicates then the Tanner graph of the associated 
PDC-LDPC code is free of length-$4$ cycles.
\end{Lem}
\begin{IEEEproof}
A length-$4$ cycle in a Tanner graph corresponds to two columns in the associated parity-check matrix
having two symbols $1$ at the same positions.
This configuration cannot occur when $S$ does not contain duplicates, that is, $S_1, S_2, \ldots, S_a$ are disjoint.
Hence, $\HH_s^T$ is free of length-$4$
cycles.
Following \eqref{eq:Hconv}, $\Hconv$ consists of copies of $\HH_s^T$ that are progressively
shifted down by one row from each other.
Therefore, the absence of duplicates in $S$ is sufficient to ensure that also $\Hconv$ is
free of length-$4$ cycles.
\end{IEEEproof}

It is well known that the absence of length-$4$ cycles in the Tanner graph of a code is a necessary condition for 
achieving good performance under belief propagation decoding.
A PDC-LDPC code with this feature can be easily designed by choosing at random the elements 
of $D_1, D_2, \ldots, D_a$ and checking that the condition imposed by Lemma \ref{lem:CyclesAbsence} is satisfied.

Provided that length-$4$ cycles are avoided, longer length cycles may obviously remain in the Tanner
graph of PDC-LDPC codes.
In particular, a length-$6$ cycle appears if and only if, given three elements $s_{i,l} \in S_i$, $s_{j,m} \in S_j$ and
$s_{k,t} \in S_k$, 
they verify the condition $s_{i,l} + s_{j,m} = s_{k,t}$.

\begin{Lem}
\label{lem:LhBound}
For a PDC-LDPC code free of length-$4$ cycles in the Tanner graph,
the syndrome former matrix $\HH_s^T$ has a number of rows
\[L_h \ge 1 + \frac{aw(w-1)}{2}. \]
\end{Lem}
\begin{IEEEproof}
The cardinality of $D_i$ is $\left|D_i\right| = w-1$, $i = 1,2,\ldots,a$.
Therefore, the number of possible sums of $j$ consecutive elements of $D_i$ is $w-j$, $j = 2,3,\ldots,w-1$.
It follows that the cardinality of $S_i$ is
\[ \left|S_i\right| = \sum_{j=1}^{w-1}(w-j) = \frac{w(w-1)}{2}, \]
with $i = 1,2,\ldots,a$, and the cardinality of $S$ is $\left|S\right| = a\left|S_i\right|$.
Since all the elements of $S$ must be different by Lemma \ref{lem:CyclesAbsence},
it follows that $\max\{S\} \ge \left|S\right|$ and $L_h \ge 1+\max\{S\} \ge 1+\left|S\right| = 1 + a\left|S_i\right|$,
where $\max\{S\}$ is the maximum in $S$.
\end{IEEEproof}

Lemma \ref{lem:LhBound} establishes a lower bound on the value of $L_h$ that can
be approached or even reached by actual codes.
In practical applications, it is important to keep $L_h$ as low as possible,
to reduce the syndrome former memory \cite{Pusane2011}.
On the other hand, we will see in the following that an increased value of $L_h$
can be the price to pay for improving other characteristics of the code, like 
its minimum distance.

\begin{Def}
We define a {\itshape uniform} PDC-LDPC code as the code having
$D_i$ ($i = 1, 2, \ldots, a$) with elements:
\begin{align}
d_{i,2m+1} & = 4^m[2(a-i)+1], \nonumber \\
d_{i,2m+2} & = 4^m(4i-2),
\label{eq:UniformPDC}
\end{align}
where $m = 0, 1, \ldots, \left\lfloor \frac{w-2}{2} \right\rfloor$, $w \ge 2$.
Note that each $D_i$ has $w-1$ elements; hence, \eqref{eq:UniformPDC}
must be used to find up to $w-1$ values for each $i$.
\end{Def}

It is easy to verify that a uniform PDC-LDPC code satisfies the hypothesis of Lemma \ref{lem:CyclesAbsence},
so its Tanner graph is free of length-$4$ cycles.
An example of $\HH_s$ is reported in \eqref{eq:UniformHs}, for a uniform PDC-LDPC code with $a=4$ and $w=4$.

\begin{figure*}[!b]
\hrulefill
\begin{equation}
\HH_s = \left[ \begin{array}{@{\hspace{1mm}}c@{\hspace{1mm}}@{\hspace{1mm}}c@{\hspace{1mm}}@{\hspace{1mm}}c@{\hspace{1mm}}@{\hspace{1mm}}c@{\hspace{1mm}}@{\hspace{1mm}}c@{\hspace{1mm}}@{\hspace{1mm}}c@{\hspace{1mm}}@{\hspace{1mm}}c@{\hspace{1mm}}@{\hspace{1mm}}c@{\hspace{1mm}}@{\hspace{1mm}}c@{\hspace{1mm}}@{\hspace{1mm}}c@{\hspace{1mm}}@{\hspace{1mm}}c@{\hspace{1mm}}@{\hspace{1mm}}c@{\hspace{1mm}}@{\hspace{1mm}}c@{\hspace{1mm}}@{\hspace{1mm}}c@{\hspace{1mm}}@{\hspace{1mm}}c@{\hspace{1mm}}@{\hspace{1mm}}c@{\hspace{1mm}}@{\hspace{1mm}}c@{\hspace{1mm}}@{\hspace{1mm}}c@{\hspace{1mm}}@{\hspace{1mm}}c@{\hspace{1mm}}@{\hspace{1mm}}c@{\hspace{1mm}}@{\hspace{1mm}}c@{\hspace{1mm}}@{\hspace{1mm}}c@{\hspace{1mm}}@{\hspace{1mm}}c@{\hspace{1mm}}@{\hspace{1mm}}c@{\hspace{1mm}}@{\hspace{1mm}}c@{\hspace{1mm}}@{\hspace{1mm}}c@{\hspace{1mm}}@{\hspace{1mm}}c@{\hspace{1mm}}@{\hspace{1mm}}c@{\hspace{1mm}}@{\hspace{1mm}}c@{\hspace{1mm}}@{\hspace{1mm}}c@{\hspace{1mm}}@{\hspace{1mm}}c@{\hspace{1mm}}@{\hspace{1mm}}c@{\hspace{1mm}}@{\hspace{1mm}}c@{\hspace{1mm}}@{\hspace{1mm}}c@{\hspace{1mm}}@{\hspace{1mm}}c@{\hspace{1mm}}@{\hspace{1mm}}c@{\hspace{1mm}}@{\hspace{1mm}}c@{\hspace{1mm}}@{\hspace{1mm}}c@{\hspace{1mm}}}
1 & 0 & 0 & 0 & 0 & 0 & 0 & 1 & 0 & 1 & 0 & 0 & 0 & 0 & 0 & 0 & 0 & 0 & 0 & 0 & 0 & 0 & 0 & 0 & 0 & 0 & 0 & 0 & 0 & 0 & 0 & 0 & 0 & 0 & 0 & 0 & 0 & 1 \\
1 & 0 & 0 & 0 & 0 & 1 & 0 & 0 & 0 & 0 & 0 & 1 & 0 & 0 & 0 & 0 & 0 & 0 & 0 & 0 & 0 & 0 & 0 & 0 & 0 & 0 & 0 & 0 & 0 & 0 & 0 & 1 & 0 & 0 & 0 & 0 & 0 & 0 \\
1 & 0 & 0 & 1 & 0 & 0 & 0 & 0 & 0 & 0 & 0 & 0 & 0 & 1 & 0 & 0 & 0 & 0 & 0 & 0 & 0 & 0 & 0 & 0 & 0 & 1 & 0 & 0 & 0 & 0 & 0 & 0 & 0 & 0 & 0 & 0 & 0 & 0 \\
1 & 1 & 0 & 0 & 0 & 0 & 0 & 0 & 0 & 0 & 0 & 0 & 0 & 0 & 0 & 1 & 0 & 0 & 0 & 1 & 0 & 0 & 0 & 0 & 0 & 0 & 0 & 0 & 0 & 0 & 0 & 0 & 0 & 0 & 0 & 0 & 0 & 0 \\
\end{array}
\right]
\tag{3}
\label{eq:UniformHs}
\end{equation}
\end{figure*}

Next we provide some results on the minimum distance of PDC-LDPC codes and uniform PDC-LDPC codes, in particular.

\begin{Pro}[\cite{Bocharova2009}]
\label{pro:Weight2DistCycles}
A code having a regular parity-check matrix with column weight $w=2$ and minimum cycle length $g$
in its Tanner graph has minimum distance $d_{\min} = g/2$.
\end{Pro}

It follows from Proposition \ref{pro:Weight2DistCycles} that a PDC-LDPC code with $w=2$ that
satisfies the hypothesis of Lemma \ref{lem:CyclesAbsence} has minimum distance $d_{\min} \ge 3$,
since length-$4$ cycles are avoided while length-$6$ cycles may still exist.
On the other hand, for uniform PDC-LDPC codes the following Theorem holds.

\begin{The}
A uniform PDC-LDPC code with $w=2$ has cycles with length $\ge 8$ in the Tanner graph
and minimum distance $d_{\min} \ge 4$.
\end{The}
\begin{IEEEproof}
Starting from the definition of uniform PDC-LDPC codes, it is easy to verify that, for any 
three elements $s_{i,l} \in S_i$, $s_{j,m} \in S_j$ and $s_{k,t} \in S_k$, 
the condition $s_{i,l} + s_{j,m} = s_{k,t}$ never occurs.
Therefore, the length of the cycles in the Tanner graph of a uniform PDC-LDPC code with $w=2$ is $\ge 8$.
It follows from Proposition \ref{pro:Weight2DistCycles} that the minimum distance is $d_{\min} \ge 4$.
\end{IEEEproof}

\begin{Exa}
Let us consider the PDC-LDPC code with $a=4$, $w=2$ and $d_{1,1}=4$, $d_{2,1}=3$, $d_{3,1}=2$, $d_{4,1}=1$.
It is easy to verify that it has minimum distance $3$.
If we instead consider the uniform PDC-LDPC code with the same parameters, it has
$d_{1,1}=7$, $d_{2,1}=5$, $d_{3,1}=3$, $d_{4,1}=1$ and it can be easily verified
that the minimum distance is increased to $4$.
On the other hand, the considered PDC-LDPC code has 
$L_h = 5$, which meets the lower bound given by Lemma \ref{lem:LhBound},
while the corresponding uniform PDC-LDPC code has $L_h = 8$.
Therefore, the improvement in the minimum distance is paid in terms of the syndrome former
memory.
\end{Exa}

\begin{Pro}
\label{pro:EvenWeight}
Regular LDPC codes with odd column weight $w$ have even minimum distance and
all their codewords have even weight.
\end{Pro}

\begin{Pro}
\label{pro:Weight3Distance4}
Regular LDPC codes with column weight $w=3$, free of length-$4$ cycles
in their Tanner graph, have minimum distance $d_{\min} \ge 4$.
\end{Pro}

\begin{The}
A uniform PDC-LDPC code with $w=3$ has minimum distance $d_{\min} \ge 6$.
\end{The}
\begin{IEEEproof}
By Proposition \ref{pro:Weight3Distance4}, a uniform PDC-LDPC code with $w=3$
has minimum distance $d_{\min} \ge 4$.
It is easy to prove that, in order to have $d_{\min} = 4$, each column in
a group of four columns summing into the all-zero vector must be included
in two distinct length-$6$ cycles involving other two columns within the 
same group.
It can be verified that, for uniform PDC-LDPC codes, this never
occurs, so $d_{\min} \ge 5$.
By Proposition \ref{pro:EvenWeight}, a uniform PDC-LDPC code with $w=3$
can only have even weight codewords, so it must be $d_{\min} \ge 6$.
\end{IEEEproof}

\begin{Exa}
Let us consider the PDC-LDPC code with $a=3$ and $w=3$, defined by:
$d_{1,1} = 5$, $d_{2,1} = 3$, $d_{3,1} = 1$, $d_{1,2} = 4$, $d_{2,2} = 7$, $d_{3,2} = 12$.
It can be easily verified that it has $d_{\min} = 4$.
In fact, $d_{2,1} + d_{3,1} = d_{1,2}$ and $d_{1,1} + d_{2,2} = d_{3,2}$,
so each column of $\HH_s^T$ is involved in two distinct length-$6$ cycles,
that is a sufficient condition for having $d_{\min} = 4$.
If we instead consider the uniform PDC-LDPC code with the same parameters,
characterized by $d_{1,1} = 5$, $d_{2,1} = 3$, $d_{3,1} = 1$, $d_{1,2} = 2$, $d_{2,2} = 6$, $d_{3,2} = 10$,
such condition is no longer satisfied and the code has $d_{\min} = 6$.
\end{Exa}

\begin{The}
A PDC-LDPC code with tail-biting termination and block length $n$ has minimum
distance $d_{\min} \le 2w$ and minimum distance multiplicity $A_{d_{\min}} \approx {a \choose 2} \cdot \frac{n}{a}$.
\end{The}
\begin{IEEEproof}
Let us consider the parity-check matrix $\HH$ of a tail-biting PDC-LDPC code 
with parameters $a$ and $w$.
Let us reorder the columns of $\HH$ by taking first those at positions $1, a+1, 2a+1, \ldots$,
then those at positions $2, a+2, 2a+2, \ldots$, and so on.
We obtain the parity-check matrix $\HH^c$ of an equivalent code, in the form of a 
single row of $a$ circulant blocks: $\HH^c = \left[ \HH^c_1| \HH^c_2| \ldots| \HH^c_a \right]$, 
each with size $n/a$ and row and column weight $w$.
For parity-check matrices in this form, it is shown in \cite{Baldi2011a} that
$d_{\min} \le 2w$ and $A_{d_{\min}} \approx {a \choose 2} \cdot \frac{n}{a}$.
\end{IEEEproof}

\section{Code examples}
\label{sec:CodeExamples}

The main advantages of uniform PDC-LDPC codes with respect to other
time-invariant LDPC convolutional codes are the guaranteed minimum 
distance and the very small memory requirements.
An important aspect is that the minimum distance value, though not 
being very high, is guaranteed independently of the code rate, which
allows designing very high rate codes with known distance.
In order to assess these benefits, we compare PDC-LDPC codes with
codes obtained through the unwrapping procedure proposed in \cite{Tanner2004},
which is known to provide very good results.

We have designed two LDPC convolutional codes with rate $R = 7/8$
by applying the method described in \cite{Tanner2004}.
With the meaning of the parameters as in \cite{Tanner2004}, these two 
codes correspond to $j = 3, k = 24, m = 73, a = 7, b = 8$ and 
$j = 4, k = 32, m = 193, a = 8, b = 81$, respectively. For these
codes, the value of $j$ also coincides with the weight of the
parity-check matrix columns.
We have chosen the smallest values of $m$ and, hence, $m_s$ that
can be found for these two pairs of $j$ and $k$. Thus, for these
parameters, the two codes have the minimum constraint length
achievable through the approach in \cite{Tanner2004}.
The values of the memory order and the constraint length are
$m_s = 71, v_s = 1728$ for the first code, and $m_s = 191, v_s = 6144$
for the second code.
The minimum distance is $6$ and $8$, respectively.
We have compared them with two uniform PDC-LDPC codes having the
same code rate, parity-check matrix column weight and minimum distance.
For the two uniform PDC-LDPC codes, the constraint length is, respectively,
$v_s = 256$ and $v_s = 624$, which are about $7$ and $10$ times less than 
those of the codes designed following \cite{Tanner2004}.

\begin{figure}
\centering
\includegraphics[keepaspectratio,width=75mm]{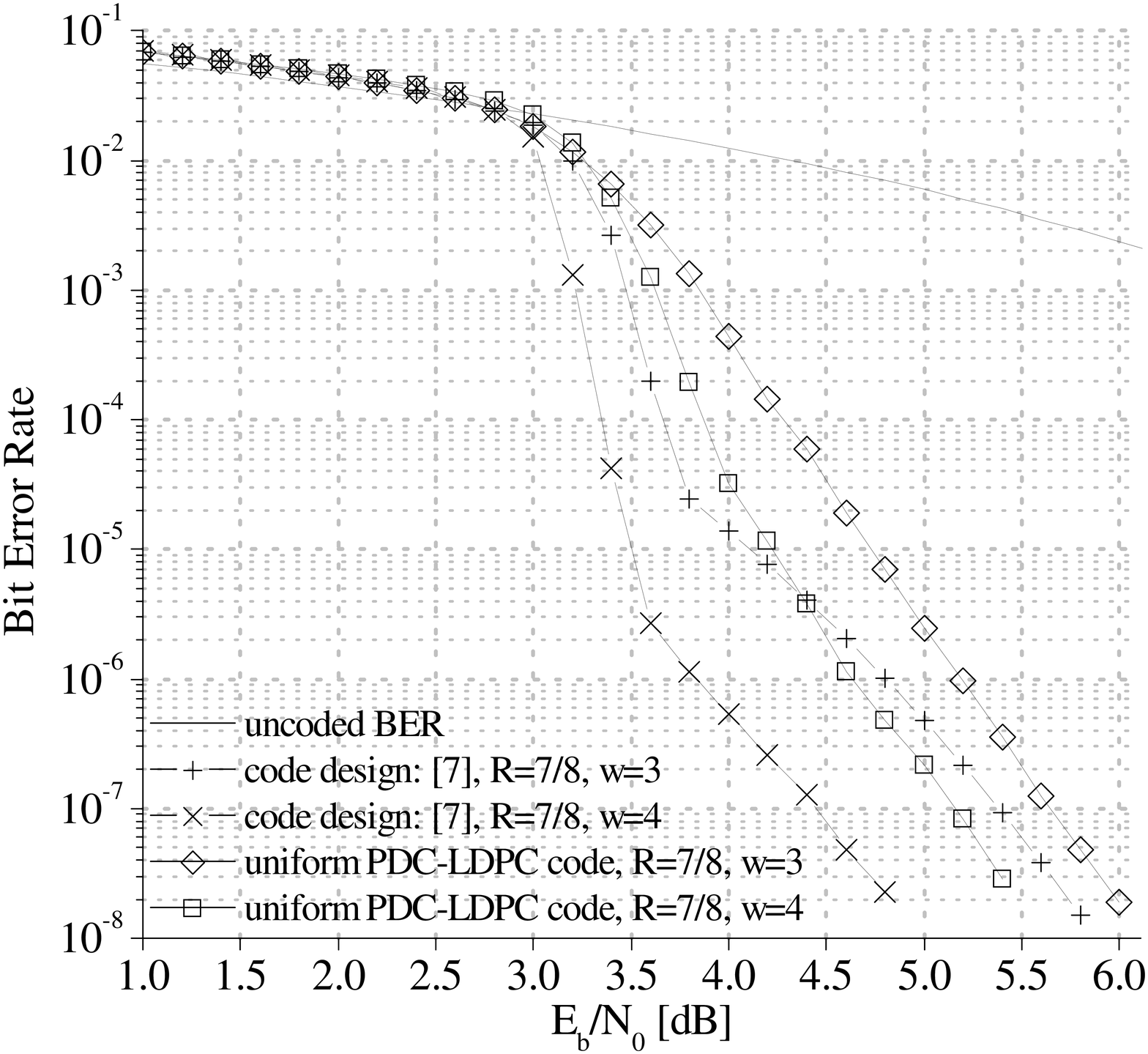}
\caption{Comparison between the solution proposed in \cite{Tanner2004}
and uniform PDC-LDPC codes for the same code rate ($R=7/8$) and column weight ($w=3,4$).}
\label{fig:Comparison}
\end{figure}

The performance of these codes has been assessed by simulating transmission 
over the additive white Gaussian noise channel, using
binary phase shift keying modulation.
The unterminated LDPC convolutional codes have been decoded by performing
the sum-product algorithm with log-likelihood ratios on sufficiently long blocks.
The simulation results, reported in Fig. \ref{fig:Comparison},
evidence that the codes from \cite{Tanner2004} have a better
waterfall performance, due to their graph properties, but then
their curves approach those of uniform PDC-LDPC codes and tend
to the same slope. 
For $w=3$, in particular, the performance gap between the two
designs becomes very small in the error floor region.
Hence, uniform PDC-LDPC codes are able to greatly reduce
the memory requirements by assuring a moderate performance loss.
This conclusion would be even more evident for higher rate codes,
while for lower code rates the approach in \cite{Tanner2004} 
achieves excellent performance.

\begin{figure}
\centering
\includegraphics[keepaspectratio,width=75mm]{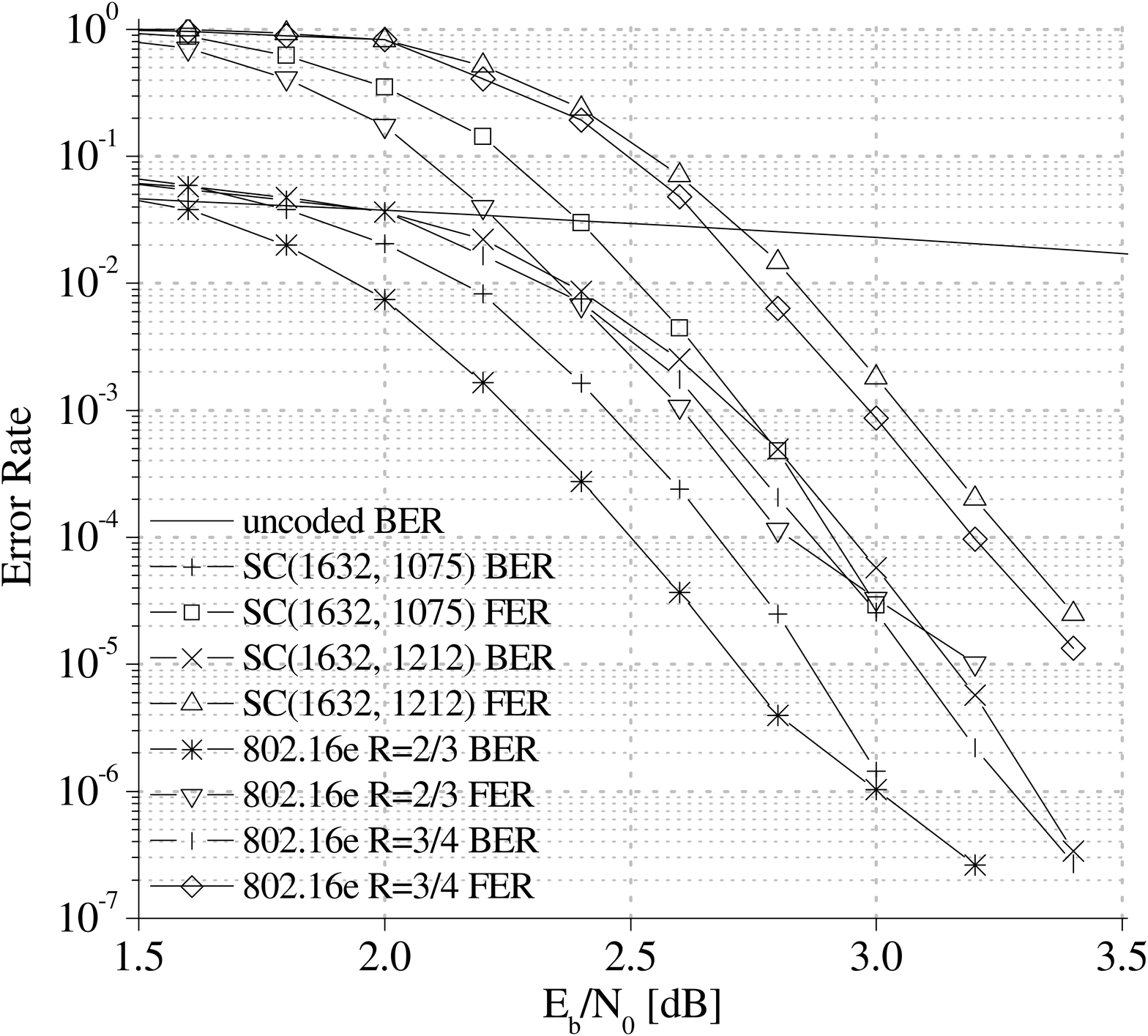}
\caption{Serially concatenated codes with $n=1632$ and rates $R \simeq 2/3, 3/4$,
compared with IEEE 802.16e standard codes.}
\label{fig:SC}
\end{figure}

Owing to their high code rate and short constraint length,
PDC-LDPC codes can be efficiently employed in concatenated schemes.
We have considered the serial concatenation (SC) of a uniform PDC-LDPC outer
code and a multiple serially-concatenated multiple-parity-check (M-SC-MPC) inner code \cite{BaldiCL2009}.
Both they are LDPC codes; hence, the overall concatenated code is
an LDPC code as well, and it has been designed in such a way as to avoid short 
cycles in its Tanner graph.
We have focused on a set of parameters used in the IEEE 802.16e standard \cite{802.16e}
and designed two SC codes with $n=1632$ and rates $R \simeq 2/3, 3/4$.
In both cases, the outer uniform PDC-LDPC code has $a=7$ and $w=2$.
Concerning termination, the $R \simeq 3/4$ code is a tail-biting code,
while the $R \simeq 2/3$ code uses a \textit{staircase} termination 
block \cite{Yang2004}, which is often used for designing systematic LDPC codes
and produces a negligible performance loss.
The simulation results are reported in Fig. \ref{fig:SC}:
The performance of the considered SC codes
is very close to that of the standard QC-LDPC codes \cite{802.16e}
in the waterfall region, and can be even better in the error floor region.

\section{Conclusion}
\label{sec:Conclusion}

We have proposed a new family of LDPC convolutional codes, named PDC-LDPC codes, 
and studied their properties.
We have defined a special class of these codes, named uniform PDC-LDPC codes,
that make the code design easier and ensure some desirable features,
like an increased minimum distance.

PDC-LDPC codes are time-invariant in $\HH$ and have two main advantages:
i) very small syndrome former memory and ii) guaranteed minimum distance,
independently of the code rate.
This allows designing good codes with high rate.

These codes have also the desirable feature that any matrix obtained as a 
segment of $\Hconv$ has full-rank. So, an efficient encoding procedure 
can be implemented, based on $\Hconv$ \cite{Felstrom1999}.
Finally, though not used in this paper, a systematic generator matrix for
uniform PDC-LDPC codes can be easily found, thus in turn enabling 
the use of very simple encoding circuits.


\end{document}